\documentclass{article}
\usepackage[utf8]{inputenc}
\usepackage[T1]{fontenc}
\usepackage{graphicx}
\usepackage{hyperref}
\usepackage{upgreek} 

\usepackage[a4paper, margin=1in]{geometry}
\usepackage{authblk}

\makeatletter
\def\@maketitle{%
  \newpage
  \null
  \vskip 2em%
  \begin{center}%
  \let \footnote \thanks
    {\LARGE\bfseries\@title \par}%
    \vskip 1.5em%
    {\large
      \lineskip .5em%
      \begin{tabular}[t]{c}%
        \@author
      \end{tabular}\par}%
    \vskip 1em%
    {\large \@date}%
  \end{center}%
  \par
  \vskip 1.5em}
\makeatother

\usepackage{cite}

\newcommand{\mum}{$\upmu{}$m}
\newcommand{\muj}{$\upmu{}$J}

\title{The mechanism of high harmonic generation\\ in liquid alcohol}
\author[1]{O. Alexander}
\author[1]{J.C.T. Barnard}
\author[1,2]{E.W. Larsen}
\author[1]{T. Avni}
\author[1]{S. Jarosch}
\author[1]{C. Ferchaud}
\author[1]{A. Gregory}
\author[1]{S. Parker}
\author[1]{G. Galinis}
\author[1]{A. Tofful}
\author[1]{D. Garratt}
\author[1]{M.R. Matthews}
\author[1]{J.P. Marangos}

\affil[1]{Quantum Optics and Laser Science Group, Blackett Laboratory, Imperial College London, London, SW7 2BW, UK }
\affil[2]{IMEC, Kapeldreef 75, 3001 Leuven, Belgium}

\begin{document}
\maketitle
\flushbottom{}
\textbf{The observation of non-perturbative harmonic emission in solids from ultrashort laser pulses \cite{GhimireNatPhys2010} sparked a wave of studies\cite{YouNatComm2017,Ghimire2019} as a  probe of charge carrier dynamics in solids under strong fields and a route to extreme ultraviolet (XUV) attosecond photonic devices\cite{GargNature2016}. High harmonic generation (HHG) in liquids\cite{LuuNatComm2018b,DiChiaraOE2009} is far less explored, despite their relevance to biological media, and the mechanism is hotly debated. Using few-cycle pulses below the breakdown threshold, we demonstrate HHG in alcohol with data showing carrier-envelope-phase-dependent XUV spectra extending to 50 eV from isopropanol. We study the mechanism of the harmonic emission through its dependence on the driving field and find it to be consistent with a strong-field recombination mechanism. This maps emitted photon energy to the electron trajectories. We explore the role of the liquid environment in scattering the trajectories and find evidence that information on electron scattering from neighbouring molecules is encoded in the harmonic spectra. Using simulations we exploit this to estimate the scattering cross section and we confirm that the cross-section in liquid isopropanol is significantly reduced compared to vapour. Our findings suggest an \textit{in situ} measurement strategy for retrieving accurate values of scattering cross sections in liquids, and also a pathway to liquid-based attosecond XUV devices.}

Bright coherent high harmonic generation is a universal phenomenon observed when a strong laser field interacts with a phase of matter, e.g. plasmas, gases, solids and liquids.  In gases this method is widely used for producing attosecond XUV pulses \cite{HentschelNature2001,PaulScience2001}  for studying the structure and dynamics of matter at electronic timescales\cite{BakerScience2006,KoturNatComm2016,PertotScience2017}.
In a real-space recombination picture, a strong laser field distorts the atomic/molecular potential or band-structure of the material \cite{crassee2017strong}. Electron wavepackets, formed by tunnelling through the distorted potentials, are accelerated and returned by the oscillating field to the hole state. The accrued kinetic energy is converted into photons ranging from XUV to soft X-ray energies. In atomic gases the highest achievable photon energy, $\hbar\omega_{max}$, can be estimated with the classical cutoff law:  $\hbar\omega_{max} = I_{pot} +3.17U_{ponder}$ where  $I_{pot}$ and $U_{ponder}$ are the ionisation potential and pondermotive energy respectively. This manifests in a cutoff energy which is linear with intensity.  In solids a momentum-space picture is used to describe strong field driven electron dynamics with a recombination component occurring between electronic bands (inter-band), and a non-linear intra-band component.  Both processes give cutoff energies which are linear with the electric field amplitude.  

In contrast the mechanism behind HHG in the liquid phase is less well understood. This is partly due to the challenges of detecting XUV from thick and absorbing cylindrical jets or spherical droplets. 
Non-perturbative harmonics\cite{DiChiaraOE2009,LuuNatComm2018b} generated in the liquid phase alone were made possible with new technology for optically flat and thin sheets of liquid  \cite{GalinisRSI2017, EkimovaSD2015}. Surprisingly these early studies of liquid phase HHG observed much lower harmonic cutoff energies than those seen in gases or bulk solids with equivalent $I_{pot}$ or bandgap. For example a study on 120 nm thick SiO$_2$ achieved 42 eV\cite{LuuNatComm2015}, while work on a 1.2-1.9 \mum{} thick sheet of ethanol achieved 25 eV\cite{LuuNatComm2018b}. 

We find that higher energy XUV emission from liquids, up to 50~eV,   does in fact occur and is made possible using a few-cycle driving field with an intensity maintained below the breakdown threshold.  Above this threshold, the overdriven regime, plasma reflection and ionisation saturation effects clamp the intensity, distort the driving pulse, reduce the harmonic flux and eliminate the highest order harmonics. Below this threshold, we find evidence that the HHG emission follows a recombination mechanism and use this to study low energy (10-50 eV) electron scattering in liquids.  

In studies on dense water vapour \cite{KurzPRA2013}, Kurz et al. showed that increasing vapour density led to a decreasing cutoff energy and attributed this to increasing electron scattering rates. A similar phenomenon is expected to occur in liquids where nearest neighbours are 2-3\,\AA\ apart, an order of magnitude smaller than the electron excursion length in the strong field (a few tens of angstroms is typical for a SWIR field generating XUV harmonics).  Gas phase electron-molecule scattering cross sections at low kinetic energy do not contain the effects of screening, correlation and exchange potentials\cite{Sinha2021,signorell2020electron,Gartman}, which are expected in dense molecular environments such as liquids.  Scattering cross sections are dependent on the electron kinetic energies, and information about these kinetic energies is contained within harmonic emission spectra. By a comparison between our experimental data and numerical simulations we advance towards extracting more accurate cross sections in liquids. 

\section*{Results}

\begin{figure}[t!]%
\begin{centering}
\includegraphics[width=1\textwidth]{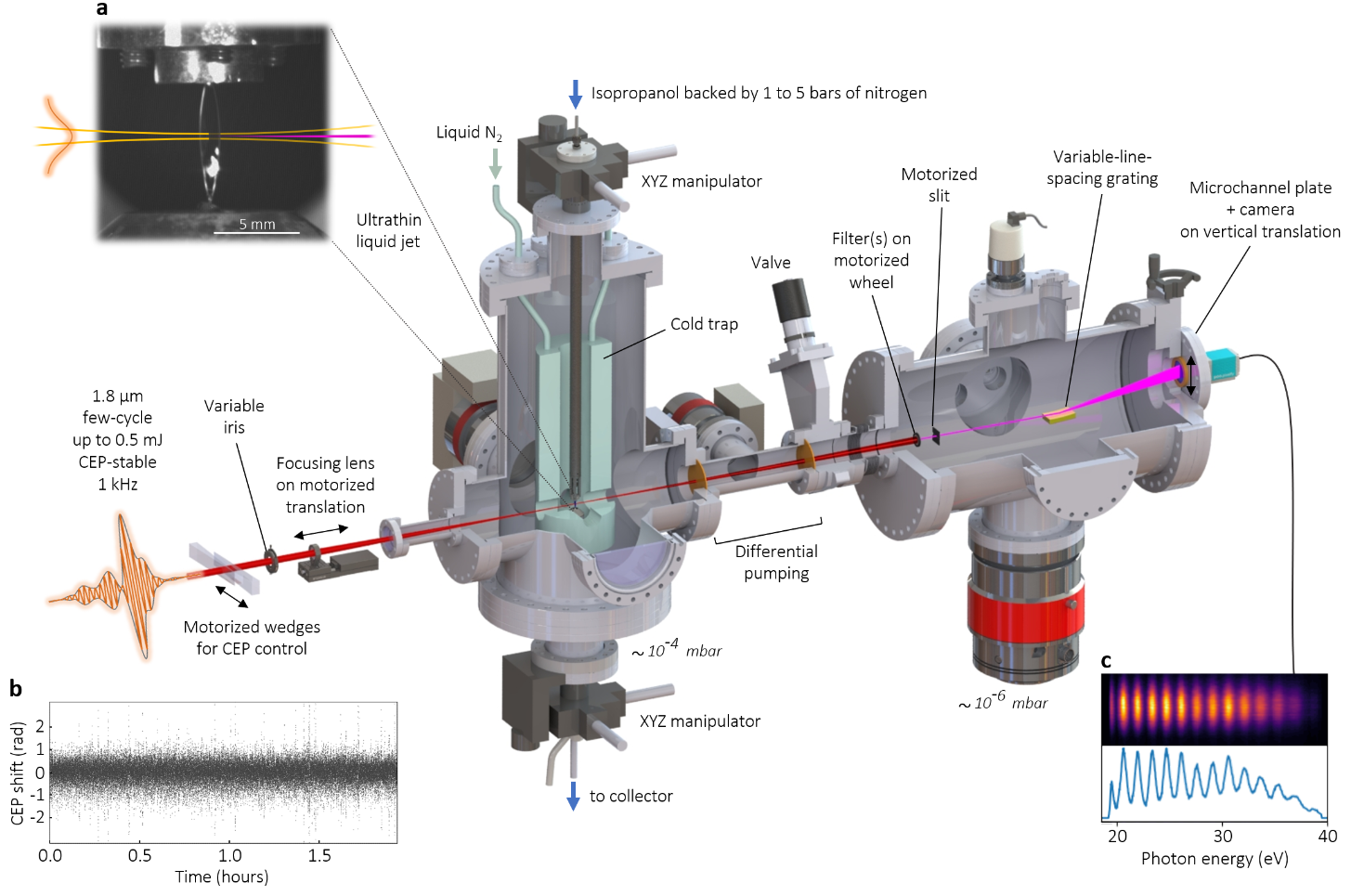}
\caption{\textbf{Experimental apparatus.} Few-cycle, CEP-controlled, linearly-polarized pulses at 1.8 \mum{}, ranging from 100 \muj{} to 350 \muj{} energy are focused into a <2 $\mu$m thick sheet of liquid running in vacuum ($10^{-4}$ mbar). Harmonic emission is detected on a flat-field XUV spectrometer consisting of a filter, slit, variable-line-spacing grating, microchannel plate and camera. \textbf{a}, In-vacuum image of the liquid jet sheet, with the laser beam position indicated. \textbf{b}, Driving-laser CEP stability over 110 minutes. \textbf{c}, Top: Spatial-spectral profile of harmonics generated in liquid phase isopropanol. Bottom: The spectrum after spatial integration.} 
\label{fig:figure1}%
\end{centering}
\end{figure}


The experiment is carried out using a thin ($\sim{}$1.5 \mum{}) flat sheet of liquid isopropanol \cite{GalinisRSI2017}  illuminated with 11--15 fs, CEP-stable, 1.8 \mum{} pulses (Figure 1).

\begin{figure}[ht!]%
\begin{centering}
\includegraphics[width=0.9\textwidth]{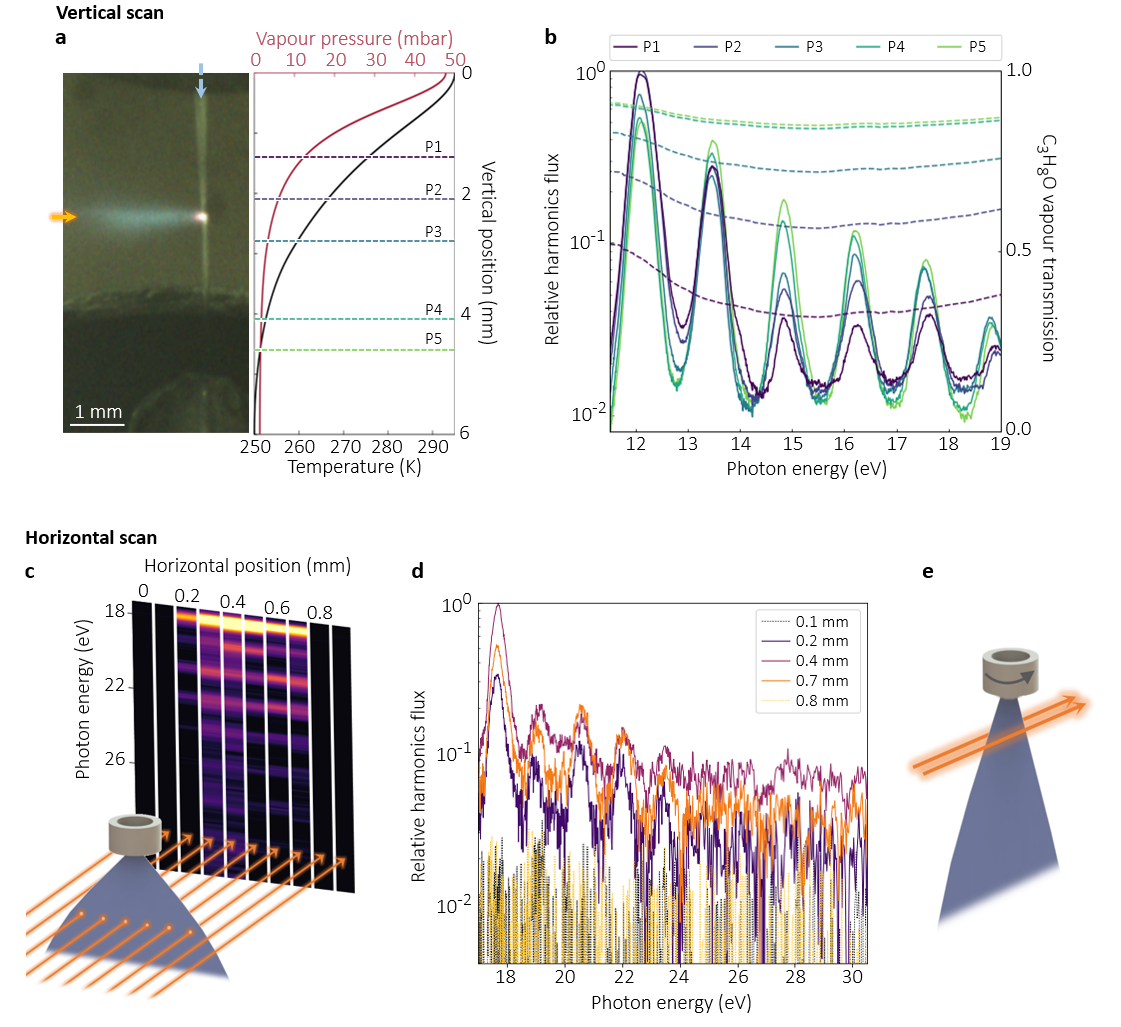}
\caption{\textbf{HHG in liquid phase.} \textbf{a}, Left: Side-view of laser incident on a jet running in vacuum, with peak intensity > 50 TW/cm$^2$.  A plasma plume is visible at the front surface of the jet. Right: Decreasing temperature (black) and vapour pressure (red) of the isopropanol due to the evaporative cooling (see Methods and Supplementary Material) occurs descending the jet. Horizontal dashed lines, P1-P5, indicate five different scan positions. The corresponding pressures are in order from P1 to P5: 12.3 mbar, 6 mbar, 3.3 mbar, 0.17 mbar, 0.16 mbar. \textbf{b}, The low-energy harmonics (12-19 eV) and isopropanol vapour transmission (dashed lines) for positions P1 to P5 shown in panel a. XUV generated at positions P5 and P1 experience significantly different vapour densities and therefore transmission. \textbf{c}, Horizontal scan of higher energy harmonics (17 to 30 eV) recorded at P1, in 11 horizontal steps of 100 $\mu$m. A sharp drop of the harmonic intensity is visible at the boundaries of the jet. \textbf{d}, Line-outs of the horizontal scan with edge positions 0.1mm and 0.8 mm (black and yellow dotted lines) showing no harmonic intensity recorded from the surrounding vapour. \textbf{e} To isolate possible emission from the surrounding vapour, the liquid target was rotated such that the laser axis is parallel to the surface.}
\label{fig:figure2}%
\end{centering}
\end{figure}

Figure 2 shows high harmonics from liquid isopropanol. The spectra are taken at different height positions along the flat liquid sheet to show the observed emission is dominantly generated in the liquid and not in the vapour surrounding the jet. The liquid interaction region ($> 1$ \mum{}) is significantly longer than the liquid-isopropanol absorption length (10-50 nm, see Supplementary Material) for the XUV. Harmonics from liquid are therefore generated near the rear surface of the liquid sheet and any harmonics generated from the vapour before the jet will not pass through.

By scanning the laser beam vertically down the jet, at the five positions shown in Figure 2a, we track the impact of the rear vapour layer on low order harmonics, where there is reliable absorption data for isopropanol\cite{KoizumiJCP1986}. Gas-based harmonics would increase with the vapour pressure but Figure 2b in fact shows a lower harmonic yield above 13 eV when the laser is positioned close to the top of the jet (P1, 15 mbar) compared to the lower positions (P4 and P5, 1 mbar). In descending the jet, the isopropanol vapour density decreases and the yield of the harmonic emission is increased due to reduced absorption of harmonics by this vapour. The change in transmission of the vapour at the different positions modifies the shape of the harmonic spectra. The greatest change is seen where isopropanol's transmission is lowest and the pressure is higher, resulting in a relative decrease in the flux. For example, the harmonic at 15 eV at P1 is less intense at P5, which corresponds to the lowest vapour transmission.

A horizontal scan of the laser pulse was performed across the upper region of the jet, where the laser passes again through 15 mbar of vapour. To identify the position on the jet, intensity of the driving field was set to above 50 TW/cm$^2$ so that a plume is clearly visible. The harmonics spectrum shows a distinct top-hat emission pattern corresponding to the physical size of the jet, with no emission detected when passing only through the vapour 200 \mum{} either side of the jet edge (Figure 2d). This was confirmed by rotating the jet sheet about the vertical axis and allowing the laser pulses to pass parallel to the sides of the jet, a few hundred microns from the liquid surface (Figure 2e). No HHG emission was observed in this configuration. 
These measurements confirm that the measured HHG emission, for both low and higher photon energies, is from liquid phase isopropanol.

\begin{figure}[t!]%
        \begin{centering}
    \includegraphics[width=1\columnwidth]{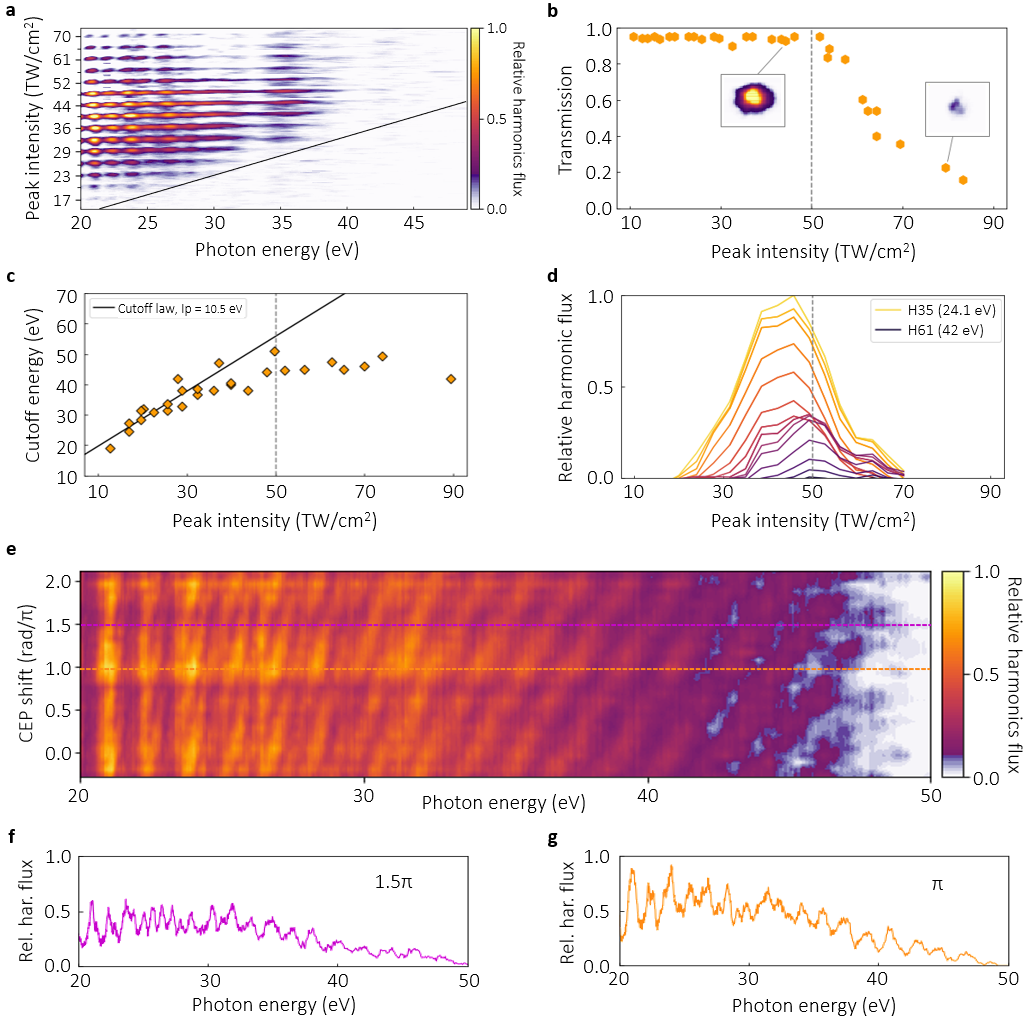}
    \caption{\textbf{Harmonics up to  50 eV.} \textbf{a}, Stacked MCP images at increasing intensities, varied using an iris. Each image show the spatial divergence in the y axis, with the discretised y axis marking the peak intensity used for the image acquisition. \textbf{b}, Laser beam transmission through the liquid sheet with increasing intensity. Insets: images of the driving beam, after the liquid sheet, at intensities below and above 50 TW/cm$^2$. Transmission drops rapidy after 53 TW/cm$^2$. \textbf{c}, Dependence of the cutoff energy on laser-field peak intensity. Classical cutoff law for Ip = 10.5 eV in solid line. \textbf{d}, Dependence of the harmonics flux on laser intensity, for harmonics 35 to 61. \textbf{e}, \textbf{f}, \textbf{g} Dependence of the harmonics spectrum on the CEP shift induced by a pair of fused silica wedges. Absolute CEP is unknown and so zero CEP is set to coincide with maximum flux.  Line-outs at the horizontal dashed lines are shown in their corresponding colours in the panels \textbf{f} and \textbf{g}.}%
    \label{fig:figure3}%
    \end{centering}
\end{figure}
We now examine the laser intensity dependence of HHG in the liquid. Figure 3a shows the  evolution of higher energy harmonics, up to 45 eV, with increasing pump intensity. The harmonic cutoff shows a broadly linear trend with laser intensity (Figure 3c), which follows the classical cutoff law up to ~35 eV, where it begins to  deviate from this. The deviation above 53 TW/cm$^2$ is explained by the sharp and rapid decrease in the laser pulse transmission at this intensity (Figure 3b). This reduction is caused by over-ionization of the liquid where the electron density exceeds the critical density for this wavelength, leading to strong reflection of the laser at the front surface \cite{lebugle2012absorption}. It also coincides with the appearance of a plume (Figure 2a) at the front surface of the jet. The plasma mirror however does not explain the damping of the cutoff (Figure 3c) nor the reduction in harmonic flux visible from 35 eV onwards in Figure 3d and suggests a second damping mechanism, such as scattering, at play. Despite this we observe a much higher cutoff than previous works, as we work below the breakdown threshold with the help of few-cycle pulse\cite{zhokhov2018optical} (see Supplementary Material) rather than multi-cycle mJ-level pulses.     
Presented in Figure 3e, f and g are harmonic spectra in optimal conditions reaching up to 50 eV, the highest reported in the liquid phase, and where the influence of the change of CEP is clearly visible across the whole spectrum. The dependence is similar to a gas phase CEP scan, with half cycle contributions appearing at the cutoff\cite{RudawskiEPJD2015}. The ellipticity dependence of the HHG is shown in the Supplementary Material and exhibits a smooth Gaussian dependence, measured below 53 TW/cm$^2$,  similar to that observed in gas phase HHG.

\begin{figure}[t!]%
\begin{centering}
\includegraphics[width=1.\textwidth]{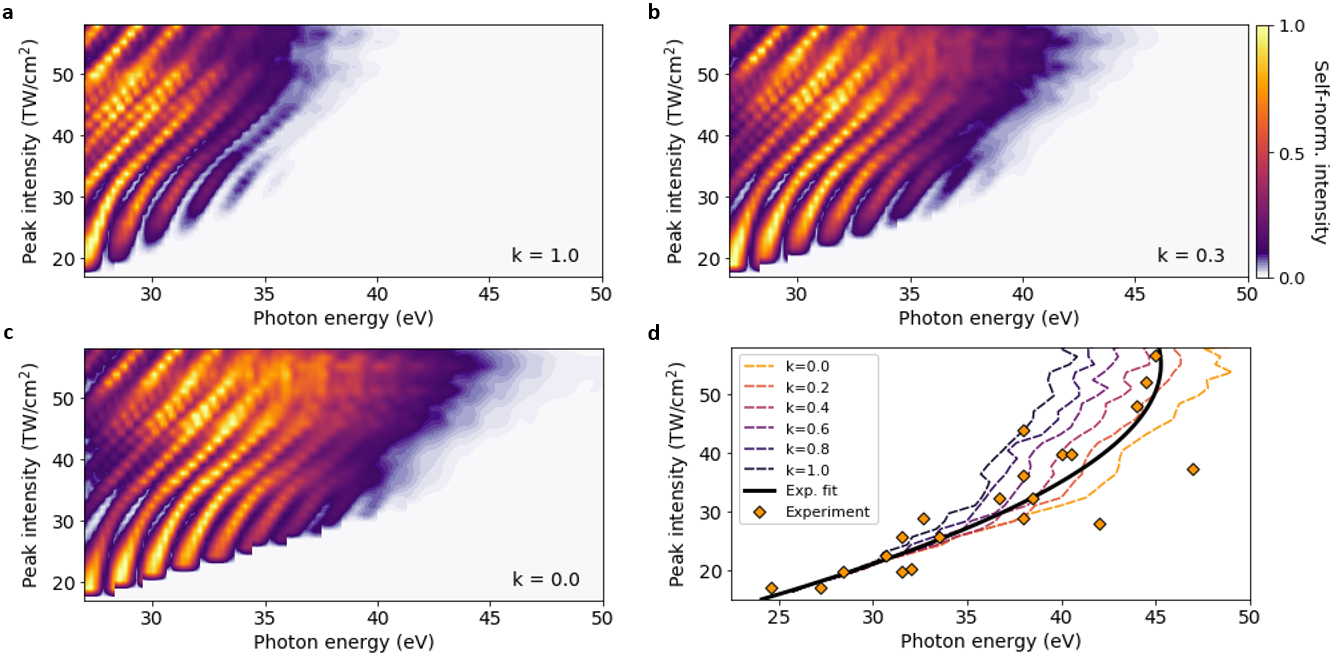}

\caption{\textbf{Numerical modelling.} Numerical modelling of high harmonic generation in isopropanol-like dense media using a semi-classical strong field approximation, corresponding to experimental conditions. A scattering probability calculation is incorporated with \textbf{a}, $k$ = 1 (see text and Methods). The higher energy trajectories are strongly damped through interactions with neighbouring molecules. \textbf{b}, $k=0.3$. \textbf{c},  $k=0$, i.e. no scattering. Harmonic intensity is self-normalised.
\textbf{d}, Harmonic cutoff energy dependence on peak intensity calculated for varying values of k (dashed coloured lines).  The experimental data is shown as orange diamonds. Measurement of harmonic cutoffs is accurate to 3 eV, due to background noise. A second order polynomial least squares-fit is shown with a solid line as a visual aid. Note that this is not a fit to $k$. }
\label{fig:figure4}%
\end{centering}
\end{figure}

Four key features are observed in our measurements: a high cutoff energy at 50 eV ; a $\pi$ periodicity of the CEP ; a linear dependence of the cutoff on intensity and a saturation effect on the harmonic spectrum above 53 TW/cm$^2$. This leads us to explore a recombination model, in which the electron wave packet following tunnelling can be considered free of the parent molecule, propagating and returning through the surrounding liquid, but with a finite probability of scattering off neighbouring molecules. We simulate the harmonic spectra based on the strong field approximation (SFA)\cite{LewensteinPRA1994} using a code that includes both an SFA kernel and full 3D field propagation, benchmarked against gas-phase HHG measurements \cite{JohnsonSciAdv2018} (see Methods).  This full propagation calculation incorporates the effect of plasma defocusing on the driving field intensity. 

To study the impact of electron scattering we apply an energy dependent filter which corresponds to a probability, for each kinetic energy of an electron, that the electron will have returned to its parent molecule without scattering.  
The probability is calculated from a simple classical model, using experimentally measured scattering cross sections \cite{bettega2011collisions} for vapourised isopropanol molecules (see Methods). For low-energy electron scattering cross sections, only vapour phase data currently exists. Therefore, in our model for the liquid phase we assume the cross section retains the same dependence on the electron kinetic energy but introduce a scaling factor, $k$, to vary the magnitude of the empirical cross-sectional data\cite{bettega2011collisions}.  $k = 1$ is equivalent to using a vapour cross-sectional area, i.e. strong scattering contribution, and $k = 0$ when no scattering is included. This is because we do not expect gas phase measurements to  accurately reflect full dynamics of electron behaviour in condensed matter. These gas phase measurements are known theoretically\cite{Sinha2021,signorell2020electron}, experimentally\cite{michaud2003cross,Gartman} and in strong field experiments \cite{schutte2016strong} to over-estimate the cross section in the condensed phase where correlation effects and charge screening events are present. This simple approach allows us to explore the effect of scattering in liquid HHG emission by encompassing the screening and correlation of the higher density medium in a single factor. 

The simulations are shown in Figure 4. Figure 4a shows calculated harmonic spectra as a function of the laser intensity with a full vapour phase scattering cross section ($k=1$). This leads to significant damping, which does not accurately model the experimental data. Figure 4b ($k=0.3$) better reflects the experimental data, with the highest energy emission still present, but dampened to the point that the harmonic cutoff appears reduced. Figure 4c ($k=0$) shows the unscattered calculation.
When we compare the experimentally determined cutoffs in Figure 4d, with the cutoffs from the simulated data for various $k$, it is clear again that weak electron scattering is present to damp harmonic generation, below the breakdown threshold (53 TW/cm$^2$), but does not significantly reduce the cutoff. Note, the modelling was performed for a pulse of 13 fs whereas experimental data associated with the CEP scans was taken with pulses ranging from 11-15fs. The shortest pulses give points which lie above the highest prediction of the modelling.  A polynomial ﬁt to second order, made to all the experimental data, is added to help compare the measured to calculated harmonic cutoff scaling with intensity.

While our elementary model is insufficient to precisely quantify $k$, we see qualitatively that the scattering cross sections of liquid isopropanol are significantly lower (approximately 3--10 times) than those previously measured in vapour. This simple but effective adaptation to SFA code gives an interesting insight into the behaviour of low energy electrons in liquids and  highlights the rich complexity of the liquid phase under strong field illumination.  A complete model that maps both the  scattering amplitude and phase of the electron onto the final emission dipole would provide further understanding.
This could be applied to measuring low energy electron scattering in liquids, which is important in radiation therapies\cite{Dosimetry,newhauser2016review}. Improved measurements, which are otherwise difficult, could improve radiation modelling and mitigating strategies for clinical outcomes.   
\section*{Conclusion}
HHG is a universal feature of light matter interaction at high intensities and can be used to extend our understanding of liquid phase dynamics.  
We have demonstrated harmonics which extend up to 50 eV in a simple alcohol. This is achieved by remaining below the breakdown threshold with the use of a few cycle pulse. The data are indicative of a recombination model in which the ionised electron is free from the parent ion. 

Furthermore, we find evidence that electron scattering does not prohibit high harmonic emission but instead plays a weak role to  dampen the emission of higher harmonics and modify the spectral shape.   This is because scattering amplitudes in the liquid phase are reduced compared to the vapour phase. We find that the HHG emission deviates from the cutoff scaling law due to increased scattering at long trajectories.

As with HHG in the solid phase, which has provided insight into charge carrier dynamics in strong fields, liquid HHG probes the transport of low energy electrons in a liquid environment. We explore one direction of future application by using a classical model to quantify scattering. This simple method signposts towards a new \textit{in situ} probe of the electron scattering amplitudes, which are otherwise challenging to access. Better determination of low energy electron scattering cross-sections will contribute to more accurate modelling of electron damage mechanisms in tissue, for example, to understand better how DNA double bonds are damaged from low energy electrons \cite{Boudaiffa2000,rezaee2017exploitation}.

\section*{Methods} 

\subsection*{Liquid jet}

The target used for the experiments in this paper was a thin and flat sheet jet produced by the fan spray nozzle presented in Galinis et al.\cite{GalinisRSI2017}. During these experiments, isopropanol was run through the nozzle, backed by 1--5 bar of compressed nitrogen. For a new nozzle, 3 bar pressure creates a sheet that is around 10 mm long and 2.5 mm wide at the middle, and 1--2 \mum{} thick over the majority of its length. The velocity of the sheet under these conditions is 18.7 m/s, which ensures that between each shot at 1 kHz repetition rate the jet has moved 1.87 cm ensuring a totally fresh target for each shot.\par

The spent liquid is captured by a collector. The entire liquid jet assembly is mounted in a vacuum chamber, and is enclosed by a liquid nitrogen filled cold trap which keeps the vacuum below $10^{-3}$ mbar whilst the jet is running. Differential pumping separates the interaction and the detection chambers (below $10^{-6}$ mbar). The cold trap has window-free ports for the laser and a side-view port for imaging and alignment of the jet. The sheet of liquid can run for up to three hours before the enclosed cold trap is saturated with condensed isopropanol and the pressure rises too high for measurements.

\subsection*{Evaporation-driven cooling of jet }

The liquid jet enters into a vacuum of below $10^{-3}$ mbar at a temperature of 20 $^\circ$C. This generates a surrounding vapour layer through evaporation, and a subsequent cooling due to the heat extraction for the phase change\cite{HeisslerNJP2014}. The evaporation driven cooling creates a temperature and pressure gradient vertically along the jet. The temperature and vapour pressure are calculated in the following manner: the number of molecules $N$ vaporizing from a surface $A$ during a time interval $dt$ is given by\cite{Wyllie,Hoyst2015}

\begin{equation}
    \frac{\delta{N}}{\delta{t}} = \frac{1}{\sqrt{2pmkbT}} \times (p_{vap} - p)A,
\end{equation}

where $m$ is the mass of one molecule, $p_{vap}$ denotes the temperature dependent vapour pressure, $p$ is the pressure of the gas phase surrounding the jet, and $T$ is the temperature of the liquid. The phase transition requires a negative evaporation enthalpy $H_{vap}$ to be added to the liquid substance, leading to the temperature $\delta{T}$.
\begin{equation}
  \delta{T}= \frac{H_{vap}}{N_{A}CM}\delta{N},
\end{equation}
where $N_{A}$ denotes the Avogadro constant, $C$ the specific heat and $M$ the mass of the remaining liquid in the volume under consideration. Thus the cooling rate of the liquid jet can be estimated. The evaporation enthalpy $H_{vap}$ is assumed to be constant over the temperature range of interest.  This calculation was checked against measured data\cite{EkimovaSD2015} (see also measurement in Supplementary Material), and good agreement was found.
The liquid-vapour system here is modelled in the Knudsen diffusion regime, in which the evaporated molecules are propagating away from the source on ballistic, collision-free, trajectories \cite{HeisslerNJP2014}, and the pressure drops as $1/r$. A correction is made to account for the  jet molecules held back due to increased collisions at the surface and a thicker vapour layer surrounds the initial surface jet before the Knudsen regime is reached.

\subsection*{Laser system}

The laser pulses are generated from a high-energy optical parametric amplifier (HE-TOPAS, Light Conversion, 40 fs, 1.8 \mum, 1.2 mJ) pumped by a commercial, but extensively customised, 1 kHz Ti:Sapphire amplifier (Red Dragon, KMLabs and Crunch Technologies, 30 fs, 800 nm, 8 mJ). The idler pulses are compressed to two optical cycles (13 fs $\pm$ 2 fs, 1.8 $\mu$m, 0.6 mJ), using an argon-filled hollow-core fibre followed by a fused silica wedge-pair compressor. The temporal profile is presented in the Supplementary Material.

The laser pulses are delivered to the interaction region by a 50 cm focal length lens. The jet assembly is mounted on 3-axis manipulator arms, which allows us to translate the jet with respect to the laser beam. An iris, situated before the lens, was used to clip the outer regions of the beam profile and therefore to vary the peak intensity of the driving field in the interaction region. The pulse energy is measured after the iris and before the lens. The ellipticity was varied using a quarter-wave plate (see Supplementary Material).

The CEP is controlled by translating one of the fused silica compressor wedges with respect to the other. A small amount of light is reflected by the wedges and is used to monitor and compensate for slow drifts of the CEP. The CEP fluctuations, of the otherwise passively stable idler beam are caused by vibrations, airflow and temperature fluctuations. These fluctuations are measured by a 2f-3f interferometer and a feedback signal is sent to a piezoelectric mirror inside the OPA\cite{JohnsonSciAdv2018}. The full scan shown in Figure 3e represents an insertion of $\sim{}$100 \mum{} of fused silica and lasted approximately 1.5 hours (See Supplementary Information). Due to the low dispersion but large difference between group velocity and phase velocity at 1.8 \mum{} and the deployment of a few-cycle pulse, no significant changes in pulse compression is caused by the wedge scan, and thus no envelope effects are observed which are otherwise common at 0.8 \mum\cite{HaworthNP2007}.

\subsection*{Laser intensity calibration}

The peak intensity is calculated using standard formulas for focusing of Gaussian beams, with corrections for reflection losses in the uncoated calcium fluoride lens and the uncoated fused silica window at 1.8 \mum, and from the liquid surface itself, as well as corrections to the focus radius when using an iris (Fourier optics). In the case of few-cycle pulse, we calculate the peak intensity based on the energy contained in the central spike, not the wings.

\[I_{peak}(w_{g})=0.94\frac{2E_p}{\tau_{p}\pi w_{g}^{2}} \qquad 
w_g=2\sqrt{(w_0)^2+(0.45\times \lambda \times 500/D)^2 } \qquad
 w_{o}=\frac{f \lambda}{w_r \pi}\]
Where f is the focal length (500 mm), $w_r$ is the waist at the lens entrance, $w_g$ is the corrected focus radius, D is the iris diameter, $E_p$ is the pulse energy measured after the iris and corrected to take into account the losses, and $\tau_{p}$ is the pulse duration at FWHM.

In our setup, for example, if the iris is closed to a diameter of 10.5 mm, the measured pulse energy is 200 \muj{} and the corrected focus radius is 113 \mum. With these parameters and losses correction, a pulse duration of 13 fs leads to a peak intensity of approximately 50 TW/cm$^2$. 

Finally our intensity calculations are further crosschecked against the CEP dependent spectral cutoff energies observed when the liquid jet is replaced with a low pressure krypton gas target (see Supplementary Material). The intensity dependence is shown to be accurately described by the classical cutoff law within 3 percent at this wavelength \cite{Shiner2013}, with little impact of phasematching conditions.  This is consistent with other values observed in the literature \cite{Bruner2018}.  We note the saturation intensity for krypton is 247 TW/cm$^2$ at 1800 nm for a 12 fs pulse duration\cite{Shiner2013}.   We were also able to quantify the beam radius from images of the beam transmitted through the jet and focused onto a camera. Intensity measurements are frequently over-estimated due to beam imperfections, and we estimate a typical error range of $\pm5$ TW/cm$^2$, from shifts in pedestal-pulse heights, error on the iris diameter measurement and on the beam waist.  Cutoff measurements are taken where XUV light falls below the MCP noise floor, with an estimated error of $\pm2.5$ eV.  

\subsection*{XUV detection system and calibration}

The spectrum is recorded using a home-built flat-field spectrometer consisting of an 250 \mum{} slit, a 1200 lines/mm variable-line-spacing flat-field grating (Hitachi 001-0437) and a cesium iodine coated 40 mm diameter microchannel plate (MCP) with a backside mounted phosphor screen (Photonis). The fluorescence of the phosphor screen is collected using a PCO Pixelfly camera. The slit and grating are fully motorized in order to optimize the spectral resolution. The MCP and camera assembly is furthermore translatable along the diffraction direction in order to record the full harmonic spectrum. The spectrometer is calibrated using the approximately linear wavelength scaling at small angles of incidence and reflection together with metallic absorption edges in aluminium, zirconium and tin in the relevant region, i.e., at 15, 25, 55 and 72 eV.

\subsection*{Scattering probability}

 The probability that an electron is not scattered along its trajectory is approximately the product of probabilities that is not scattered for small time steps, $\Delta t$ between its birth and recombination. This becomes exact when the number of steps, $N$, approaches infinite such that $\Delta t \rightarrow dt$.
\[
     p_{\rm{n.s.}} = \prod_{i=1}^{N} 1-p_i\Delta t
 \] \[
     p_i = p_{\rm{scat}}(t_0 + i\Delta t)
 \]  \[
     N = \frac{t_{\rm{rec}} - t_0}{\Delta t}
 \]
 where $p_{\rm{scat}}(t)$ is the probability of scattering per unit time, at time $t$; and $t_0$ and $t_{\rm{rec}}$ are the birth and recombination times of the electron trajectory respectively.
 Expanding in orders of $\Delta t$
  \[p_{\rm{n.s.}} = 1 + \sum_{i=1}^{N} \left(-\frac{1}{n!}\sum_{j=1}^{N}p_j\Delta t\right)^i\\
     = e^{-\langle n \rangle} \]
 
 where
 \begin{equation}
\langle n\rangle = \sum_{j=1}^{N}p_j\Delta t 
 \end{equation}
 is the average number of collisions that would occur for the electron following its classical, scattering-free trajectory. For infinite $N$,
 \begin{equation}
     \langle n\rangle = \int_{t_0}^{t_{\rm{rec}}}p_{\rm{scat}}(t)dt
 \end{equation}
 The problem is therefore reduced to determining and integrating the probability of the electron scattering per unit time.

 In this billiard-ball model, the probability of scattering is given by the probability of encountering a molecule along its path. Considering an infinitesimal time step,
 \begin{equation}
     p_{\rm{scat}}(t)dt = \rho_N A v(t)dt,
\end{equation}
where $\rho_N$ is the number density of the molecules and $A$ is their cross-sectional area. For the cross-sectional areas we use empirical values measured in electron scattering experiments \cite{bettega2011collisions} multiplied by a factor of $k$ in order to encompass the multiple scattering, correlation, exchange and screening effects due to the density of the liquid\cite{boyle2016ab},
 \begin{equation}
     A = \sigma' = k \sigma_{emp.}.
\end{equation}
Finally we have:
 \begin{equation}  p_{\rm{n.s.}}= e^{- \int_{t_0}^{t_{\rm{rec}}}p_{\rm{scat}}(t)dt}
     = e^{- \int_{t_0}^{t_{\rm{rec}}} \rho_N k A v(t)dt}
     = e^{- \int_{t_0}^{t_{\rm{rec}}} \rho_N  k \sigma_{emp.} v(t)dt}
     \end{equation}
These effects arise due to the quantum nature of the electron, and tend to decrease the effective scattering cross section. 
As discussed in the Results section, the value of this factor was determined by comparing the simulated harmonics cutoff to the cutoff observed in the experimental data. 

\subsection*{Numerical simulations}

The jet is modelled as optically flat with a thickness of 2 \mum{} in accordance with interferometrically-measured values \cite{GalinisRSI2017}, and isopropanol is treated as a dense gas (290 bar) with an abrupt decay from liquid pressure to vapour pressure. We set the gas density to be 290 bar, to be consistent with the number density of isopropanol at 273 K (see Supplementary Material). 

For initial conditions a transform limited spatio-temporally Gaussian pulse, matching the measured pulse parameters, is used: a 13  fs pulse with a $1/e^2$ beam diameter of 260 \mum. 

The laser propagates using the forward Maxwell equation in three dimensions with cylindrical symmetry (dispersion, paraxial diffraction, third-order nonlinearity, plasma dephasing and absorption included). Instantaneous ionization rates are calculated using the ADK formula\cite{ammosov1986tunnel}, and integration performed using a preconditioned Runge-Kutta method with adaptive step-sizing. Dispersion and diffraction are applied in the frequency domain while nonlinear effects are applied in the real-space domain (Split-step method). 
A standard hydrogen wavefunction with isopropanol's ionisation potential is used, as a full treatment of molecules as emitters and electron-scattering targets is beyond the scope of this work.

For harmonic generation, the simulated laser field is resampled to a 1 \mum{} spatial grid, and the single-atom HHG dipole computed at every point using the strong field approximation with stationary phase approximations used for the momentum and birth time integrals. The scattering is applied to the harmonics before they are propagated to the end of the target region.

Because the scattering cross-section increases substantially with trajectory time, long trajectories are expected to contribute minimally to the macroscopic harmonics. As such, temporal apodization is used to remove the contribution of the long trajectories and reduce computation time. 
The macroscopic field is then obtained by spatially integrating the HHG single-atom dipoles, accounting for paraxial diffraction, the local medium density, absorption and dispersion.

\bibliography{ReducedReferences}{}
\bibliographystyle{naturemag}

\section*{Acknowledgements} 
This research was supported by Engineering and Physical Sciences Research Council (EPSRC EP/N018680/1 and EP/R019509/1),
Defence Science and Technology Laboratory DSTL/ (MURI EP/N018680/1) and Horizon 2020 Framework Programme (H2020)
(Marie Sklodowska-Curie, 641272). MM acknowledges the support of The Royal Society, with URF/R1/191759. We gratefully acknowledge technical assistance from C. O'Donovan, useful discussions with Misha Ivanov and John Tisch, and important contributions to the development of the liquid jet from Roland Smith. 

\section*{Author contributions statement}
OA, JB, EL, TA, SJ, AG, SP, GG, AT, MM and JM designed and constructed the experimental apparatus, electronics, liquid jet, chambers and optical systems. OA, JB, EL, TA, SJ, CF, DG and MM generated the few-cycle laser pulses. OA, JB, EL,  SJ, CF and MM recorded and analysed the data. TA and MM worked on the numerical simulations. OA, JB, TA and MM worked on the electron scattering model. OA, JB, EL, TA, SJ, CF, MM and JM interpreted data. OA, JB, EL, TA, CF, MM and JM wrote the paper. JM conceived the experiment.

\section*{Competing interests}
The authors declare no competing interests.
\end{document}